\documentclass[journal]{IEEEtran}
\usepackage{cite}
\usepackage{amsmath,amssymb,amsfonts}
\usepackage{algorithmic}
\usepackage{graphicx} 
\usepackage{epstopdf}
\usepackage{epsfig}
\usepackage{textcomp}
\usepackage{xcolor}
\usepackage{algorithm}  
\usepackage{textcomp}
\usepackage{xcolor}
\usepackage{bm}
\usepackage{setspace}
\ifCLASSINFOpdf
 
\else
 
\fi

\begin{document}

\title{Antenna Positioning and Beamforming Design for Fluid Antenna-Assisted Multi-user Downlink Communications }

\author{Haoran~Qin,~Wen~Chen,~\IEEEmembership{Senior~Member,~IEEE},~Zhendong~Li,~Qingqing~Wu,~\IEEEmembership{Senior~Member,~IEEE},~Nan~Cheng,~\IEEEmembership{Senior~Member,~IEEE},~and~Fangjiong~Chen,~\IEEEmembership{Member,~IEEE}

\thanks{H. Qin, W. Chen, Z. Li and Q. Wu are with the Department of Electronic Engineering, Shanghai Jiao Tong University, Shanghai 200240, China (e-mail: haoranqin@sjtu.edu.cn; wenchen@sjtu.edu.cn; lizhendong@sjtu.edu.cn).  

N. Cheng is with the State Key Laboratory of Integrated Services Networks, Xidian University, Xi’an 710071, China (e-mail: nancheng@xidian.edu.cn).

F. Chen is with the School of Electronic and Information Engineering, South China University of Technology, Guangzhou 510640, China (e-mail: eefjchen@scut.edu.cn).

(\emph{Corresponding author: Wen Chen.}).

}
}

\maketitle

\begin{abstract}
This paper investigates a multiple input single output (MISO) downlink communication system in which users are equipped with fluid antennas (FAs). First, we adopt a field-response based channel model to characterize the downlink channel with respect to FAs’ positions. Then, we aim to minimize the total transmit power by jointly optimizing the FAs’ positions and beamforming matrix. To solve the resulting non-convex problem, we employ an alternating optimization (AO) algorithm based on penalty method and successive convex approximation (SCA) to obtain a sub-optimal solution. Numerical results demonstrate that the FA-assisted communication system performs better than conventional fixed position antennas system.
\end{abstract}


\IEEEpeerreviewmaketitle

\section{Introduction}
Fueled by the exponential proliferation of wireless applications, an escalating imperative arises for enhanced capacity within forthcoming sixth-generation (6G). In pursuit of this objective, multiple-input multiple-output (MIMO)  has been envisioned as a pivotal enabling technology. MIMO systems have the capability to facilitate concurrent transmissions of multiple data streams by leveraging the novel degrees of freedom (DoFs) within the spatial domain. They have a wide range of application scenarios in wireless communication and also combine well with other promising technologies\cite{zhang2023intelligent,zhu2023mimo,OFDMA}. Nonetheless, due to the stationary deployment of antennas in conventional MIMO systems, they are subject to random and uncontrolled channel fading, hindering the full exploitation of channel variations in the spatial field. 

To overcome these inherent limitations, a novel antenna system, namely fluid antenna system (FAS) was proposed in \cite{FAS}, which allows the antenna position to be shaped freely in a given region. Many studies currently have arisen to investigate the significant performance of FAS. It was initially validated that FAS can achieve a lower outage probability than a multi-antenna fixed position system, even with constraint space, particularly when the number of available candidate ports is large enough\cite{FAS}. In \cite{FAScapacity}, the authors demonstrated that the substantial capacity gain of FAS and analyzed how the capacity varies with different parameters.
As FAS can leverage the spatial moments which interference suffer the deep fade to maximize the desired signal's strength, the authors in \cite{FAMA} investigated the ability of FAS to facilitate multi-user communication and characterized the multiplexing gain of fluid antenna multiple access (FAMA).

With the aforementioned advantages, the application of FAs in the realm of wireless communication has garnered significant interest\cite{zhu2023challenge}. For instance, the authors in\cite{ma2023mimo} proposed a new FA-enabled MIMO communication system and characterized its channel capacity with respect to antennas' positions. While \cite{uplinkMABS} explored the base station (BS) equipped with several FAs for boosting multi-user uplink communication performances, \cite{discrete} tackled a more practical problem by modelling the motion of the FAs as discrete movements. However, the majority of current researches focus on the deployment of FAs at the BS end, while research into the implementation of FAs at the user is still in an early stage.

Inspired by the preceding discussion, this paper investigates joint design of beamforming and the position of FAs for a novel FA-assisted downlink multi-user communication system, in which the BS with fixed position antenna (FPA) transmits data to multiple single FA users. In particular, the FA positioning at different users and transmit beamforming at the BS are jointly optimized to minimize the BS transmit power, as well as satisfy the minimum signal-to-interference-plus-noise ratio (SINR) and the limited moving region constraints. Since the resulting problem is non-convex with highly coupled variables, an alternating optimisation (AO) algorithm based on successive convex approximation (SCA) and penalty method is proposed to obtain a sub-optimal solution. 

\section{System Model and Problem Formulation}
As shown in Fig. 1, a BS equipped with total $N$ FPAs serves $K$ users. Each user is equipped with a single FA moving in the local rectangular region ${{\cal C}_k}$, i.e., ${{\cal C}_k}=\left[ {x_k^{\min },x_k^{\max }} \right] \times [y_k^{\min } \times y_k^{\max }],1 \le k \le K$. We describe its position as ${{\bf{u}}_k} = {\left[ {{x_k},{y_k}} \right]^T} \in {{\cal C}_k}$, $1 \le k \le K$  by establishing a 2D local coordinate system. Similarly, the position of the
\begin{figure}
	\includegraphics[width=9cm, height=5cm]{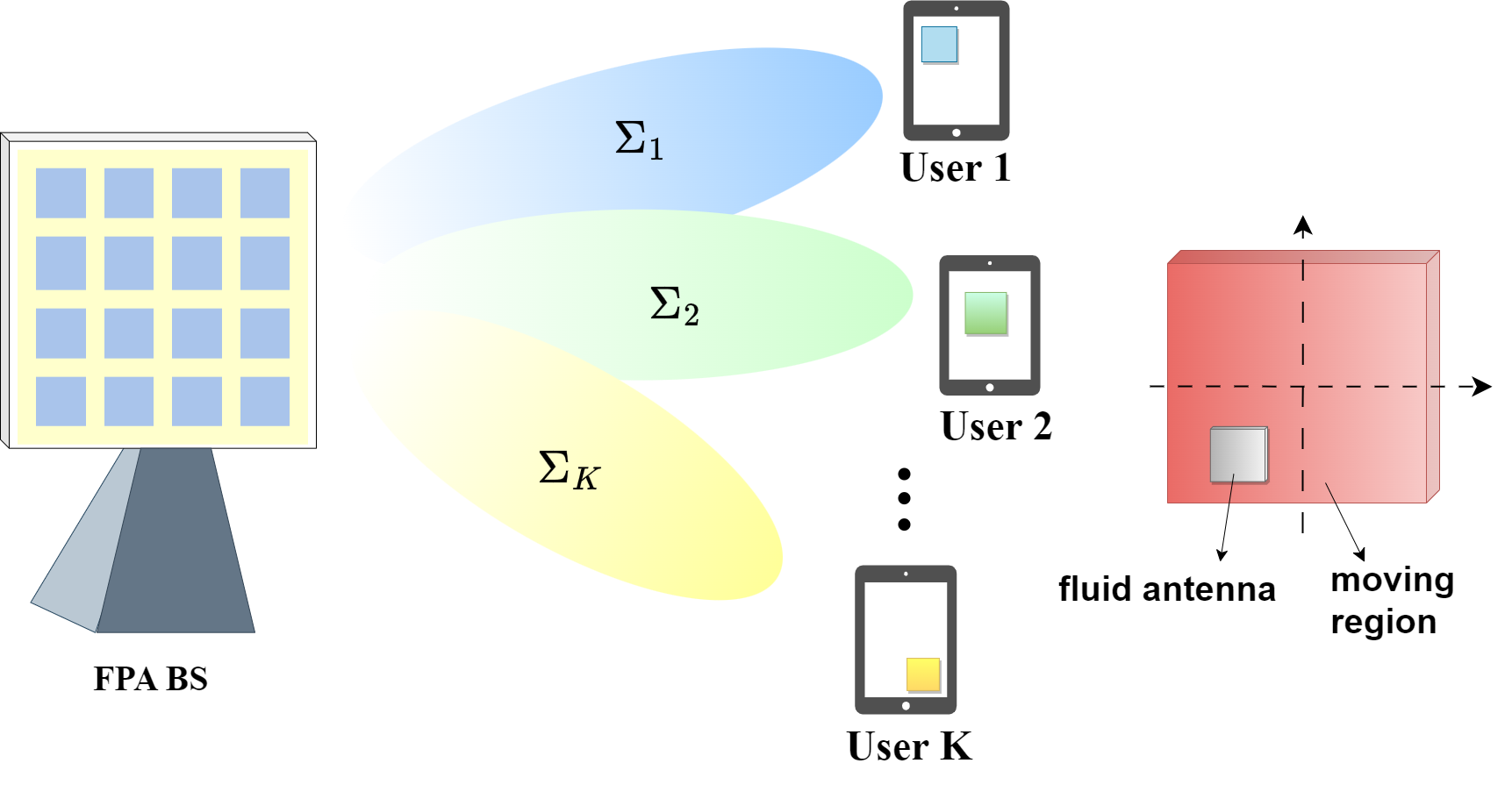}
	\caption{The FA-assisted multi-user downlink communication system.}
	\label{Fig1}
\end{figure}
 $n$-th FPA at the BS can be represented as ${{\bf{v}}_n} = {\left[ {{x_n},{y_n}} \right]^T}$, $1 \le n \le N$. During the downlink transmission from the BS to users, the received signal of the user $k$ can be expressed as 
\begin{equation}
    \begin{aligned}
        {{\bf{y}}_k}={{\bf{h}}_k}{({{\bf{u}}_k})^H}{\bf{Ws}} + {{\bf{n}}_k},
    \end{aligned}
\end{equation}
 where ${\mathbf{W}}=\left[ {{{\mathbf{w}}_1},{{\mathbf{w}}_2},\cdots,{{\mathbf{w}}_K}}\right] \in {\mathbb{C}^{N \times K}}$ is the beamforming matix at the BS, and ${{\mathbf{h}}_k}({{\mathbf{u}}_k}) \in {\mathbb{C}^{N \times 1}}$ denotes the channel vector between the BS and user $k$. ${\bf{s}} = {\left[ {{s_1},{s_2}, \cdots {s_K}} \right]^T} \in {{\mathbb C}^{K \times 1}}$ represents the transmitted signals of the users with normalized power, i.e., $\mathbb{E}\left( {{\mathbf{s}}{{\mathbf{s}}^H}} \right) = {{\mathbf{I}}_K}$. 
${{\mathbf{n}}_k} = {[{n_1},{n_2}, \cdots ,{n_N}]^T} \sim \mathcal{C}\mathcal{N}\left( {{\mathbf{0}},{\sigma ^2}{{\mathbf{I}}_N}} \right)$ is the additive white Gaussian noise (AWGN) at the user $k$ with the average noise power ${\sigma ^2}$.

For simplicity, we focus on one quasi-static fading block and the far field-response based channel model is adopted. Consequently, the angle-of-departure (AoD), the angle-of-arrival (AoA), and the amplitude of the complex coefficient for each channel path remain constant regardless of the varying positions of the FAs \cite{zhu2022modeling}.
We assume that the channal from the BS to the user $k$ has $L_k^t$ transmit paths and $L_k^r$ receive paths. 
For user $k$, the signal propagation difference for the $i$-th receive path between the $k$-th user's FA position ${{\bf{u}}_k}$ and original point ${{\mathbf{o}}_r} = {\left[ {0,0} \right]^T}$ can be described as $\rho _{k,i}^r\left( {{\bf{u}}_k} \right) = {x_k}\sin \theta _{k,i}^r\cos \phi _{k,i}^r + {y_k}\cos \theta _{k,i}^r$, where $\theta _{k,i}^r $ and $\phi _{k,i}^r $ are denoted as elevation and azimuth AoAs respectively. Thus its corresponding phase difference is $\frac{{2\pi }}{\lambda }\rho _{k,i}^r\left( {{\bf{u}}_k} \right)$, where $\lambda $ is the carrier wavelength. Accordingly, we can obtain the receive field response vector (FRV) of the user $k$ as following
	\begin{equation}
		\begin{aligned}
			{{\mathbf{f}}_k}\left( {{{\mathbf{u}}_k}} \right) = {\left[ {{e^{j\frac{{2\pi }}{\lambda }\rho _{k,1}^r({{\mathbf{u}}_k})}},{e^{j\frac{{2\pi }}{\lambda }\rho _{k,2}^r({{\mathbf{u}}_k})}}, \ldots ,{e^{j\frac{{2\pi }}{\lambda }\rho _{k,Lr}^r({{\mathbf{u}}_k})}}} \right]^T}.
		\end{aligned}
	\end{equation}
Similarly, the transmit FRV from the $n$-th FPA to the $k$-th user is given by
\begin{equation}
    \begin{aligned}
        {{\bf{g}}_{k,n}} = {\left[ {{e^{j\frac{{2\pi }}{\lambda }\rho _{k,1}^t({{\bf{v}}_n})}},{e^{j\frac{{2\pi }}{\lambda }\rho _{k,2}^t({{\bf{v}}_n})}}, \ldots ,{e^{j\frac{{2\pi }}{\lambda }\rho _{k,L_k^t}^t({{\bf{v}}_n})}}} \right]^T}, 
    \end{aligned}
\end{equation}where  $\rho _{k,j}^t\left( {{\bf{v}}_n} \right) = {x_n}\sin \theta _{k,j}^t\cos \phi _{k,j}^t + {y_n}\cos \theta _{k,j}^t$, $\theta _{k,j}^t$ and $\phi _{k,j}^t$ are the elevation and azimuth AoDs respectively. Moreover, we define a path response matrix (PRM) ${{\mathbf{\Sigma }}_k} \in {\mathbb{C}^{{L_r} \times {L_t}}}$, and element ${{\mathbf{\Sigma }}_k}\left[ {m,n} \right]$ represents the channel response between the BS origin and the receiving origin, where the signal departures from the $n$-th transmit path and is received at the $m$-th receive path. Thus, we describe the channel from the BS to the $k$-th receiver as following
\begin{equation}
    \begin{aligned}
        {{\mathbf{h}}_k}({{\mathbf{u}}_k}) = {\left( {{{\mathbf{f}}_k}{{\left( {{{\mathbf{u}}_k}} \right)}^H}{{\mathbf{\Sigma }}_k}{{\mathbf{G}}_k}} \right)^T} ,
    \end{aligned}
\end{equation} where ${{\mathbf{G}}_k} = \left[ {{{\mathbf{g}}_{k,1}}, {{\mathbf{g}}_{k,2}}, \cdots , {{\mathbf{g}}_{k,N}}} \right]$ is the field-response matrix (FRM) at the BS. Therefore, the receive SINR of the user $k$ can be described as

\begin{small}
\begin{equation}
    \begin{aligned}
        {\gamma _k} = \frac{{{{\left| {{{\mathbf{h}}_k}{{\left( {{{\mathbf{u}}_k}} \right)}^H}{{\mathbf{w}}_k}} \right|}^2}}}{{\sum\limits_{q = 1,q \ne k}^K {\left| {{{\mathbf{h}}_k}{{\left( {{{\mathbf{u}}_k}} \right)}^H}{{\mathbf{w}}_q}} \right| + {\sigma ^2}} }},\forall k.
    \end{aligned}
\end{equation}
\end{small}In this paper, we aim to minimize the transmit power of the BS by jointly optimizing the beamforming vector and the position of each user's FA. Thus the corresponding optimization problem can be expressed as
\begin{subequations}
		\begin{align}
			\left( {{\textrm{P0}}} \right){\rm{~~}}&\mathop {\min}\limits_{\left\{ {{{\bf{w}}_k},{{\bf{u}}_k}} \right\}} {\rm{~~}}\sum\limits_{k = 1}^K {{{\left\| {{{\bf{w}}_k}} \right\|}^2}},\nonumber\\ 
			\rm{s.t.}\quad&\frac{{{{\left| {{{\bf{h}}_k}{{\left( {{{\bf{u}}_k}} \right)}^H} 
								{{\bf{w}}_k}}\right|}^2}}}{{\sum\limits_{q = 1,q \ne k}^K {{{\left| {{{\bf{h}}_k}{{\left( {{{\bf{u}}_k}} \right)}^H}{{\bf{w}}_k}} \right|}^2}}  + {\sigma ^2}}} \ge {\gamma _k},\forall k,\\
			&{{\bf{u}}_k} \in {{\cal C}_k}, \forall k,
		\end{align}
	\end{subequations}where constraint (6a) represents the minimum SINR requirement of each user and constraint (6b) is the limited moving region constraint of FAs. Evidently, given the non-convex nature of both constraint (6a) and the objective function, this problem is inherently non-convex and poses a challenge for direct solution.
\section{Joint Antenna Position and Beamforming Design Algorithm}

In this section, the AO algorithm based on penalty method is proposed to solve (P0). To be specific, we decompose the penalized problem into three sub-problems and alternately optimize them until the total objective function achieving convergence. In addition, the penalty factor is gradually updated until all constraints are approximately satisfied. To this end, we first introduce auxiliary variables and the original problem can be equivalently transformed into (P1) as follows

\begin{small}
\begin{subequations}
    \begin{align}
        \left(  {{\textrm{P1}}} \right){\rm{~~}}&\mathop {\min }\limits_{\left\{ {{{\bf{w}}_k},{{\bf{u}}_k}} \right\}} {\rm{~~}}\sum\limits_{k = 1}^K {{{\left\| {{{\bf{w}}_k}} \right\|}^2}},\nonumber \\
        \rm{s.t.}\quad&\frac{{{{\left| {{t_{k,k}}} \right|}^2}}}{{\sum\limits_{q = 1,q \ne k}^K {{{\left| {{t_{k,q}}} \right|}^2} + {\sigma ^2}} }} \ge {\gamma _k},\forall k,\\
        &{t_{k,q}} = {{\bf{h}}_k}{({{\bf{u}}_k})^H}{{\bf{w}}_q},\forall k,q,\\
        &{\bf{u}}_k \in {{\cal C}_k}, \forall k.
    \end{align}
\end{subequations}
\end{small}The variables $\left\{ {{{\bf{w}}_k}} \right\}$, $\left\{ {{{\bf{u}}_k}} \right\}$ are coupled with each other in newly introduced equality constraints. To address this issue, we integrate these constraints into the objective function based on the penalty function approach. In particular, we formulate the following optimization problem by transforming these equality constraints into quadratic functions and subsequently incorporating them into the objective function as penalty terms
\begin{subequations}
    \begin{align}
        \left({{\textrm{P2}}}\right){\rm{~~}}&\mathop {\min }\limits_{\left\{ {{{\bf{w}}_k},{{\bf{u}}_k},{t_{k,q}}} \right\}} {\rm{  }}\sum\limits_{k = 1}^K {{{\left\| {{{\bf{w}}_k}} \right\|}^2}} \nonumber\\&+ \frac{1}{{2\rho }}\left( {\sum\limits_{k = 1}^K {\sum\limits_{q = 1}^K {{{\left| {{{\bf{h}}_k}{{({{\bf{u}}_k})}^H}{{\bf{w}}_q} - {t_{k,q}}} \right|}^2}} } } \right),\nonumber\\
        \rm{s.t.}\quad &\textrm {(7a), (7c)}, 
    \end{align}
\end{subequations}
where $\rho  > 0$ represents the penalty factor employed to penalize the deviation from equality constraints in (P1). It is noteworthy that even though the equality constraints are relaxed in (P2), as $\rho \to 0$, the solution derived from solving (P2) typically adheres to the constraints in (P1).

\subsection{Alternating Optimization Algorithm for Solving (P2)}

\newcounter{my1}
\begin{figure*}[!t]
	\normalsize
	\setcounter{my1}{\value{equation}}
	\setcounter{equation}{15}
	\small
        \begin{equation}
		  \begin{aligned}
                {\mathop{\rm g}\nolimits} \left( {{{\bf{u}}_k}} \right) = \sum\limits_{q = 1}^K {\left( {{{\bf{C}}_{k,q}}\left( {1,1} \right) + {{\bf{C}}_{k,q}}\left( {2,2} \right) +  \cdots  + {{\bf{C}}_{k,q}}\left( {{L_r},{L_r}} \right) + \sum\limits_{i = 1}^{{L_r} - 1} {\sum\limits_{j = i + 1}^{{L_r}} {{f_1}(i,j,k,q)} }  - \sum\limits_{l = 1}^{{L_r}} {{f_2}(l,k,q)}  + {{\left| {{t_{k,q}}} \right|}^2}} \right)} 
		  \end{aligned}
	\end{equation}
       \small
        \begin{equation}
            \begin{aligned}
                \frac{{{\partial ^2}{\mathop{\rm g}\nolimits} }}{{\partial {x_k}^2}} = &\frac{{4{\pi ^2}}}{{{\lambda ^2}}}\sum\limits_{q = 1}^K {\sum\limits_{L = 1}^{{L_r}} {{f_2}} \left( l,k,q\right){{\sin }^2}\theta _{k,l}^r{{\cos }^2}\phi _{k,l}^r} - \frac{{4{\pi ^2}}}{{{\lambda ^2}}}\sum\limits_{q = 1}^K {\sum\limits_{i = 1}^{{{L_r}} - 1} {\sum\limits_{j = i + 1}^{{L_r}} {{f_1}(i,j,k,q){{\left( {\sin \theta _{k,i}^r\cos \phi _{k,i}^r - \sin \theta _{k,j}^r\cos \phi _{k,j}^r} \right)}^2}} } } 
            \end{aligned}
        \end{equation}
        \small
        \begin{equation}
            \begin{aligned}
                \frac{{{\partial ^2}{\mathop{\rm g}\nolimits} }}{{\partial {y_k}^2}} = \frac{{4{\pi ^2}}}{{{\lambda ^2}}}\sum\limits_{q = 1}^K {\sum\limits_{l = 1}^{{L_r}} {{f_2}} \left( l,k,q\right){{\cos }^2}\theta _{k,l}^r}  - \frac{{4{\pi ^2}}}{{{\lambda ^2}}}\sum\limits_{q = 1}^K {\sum\limits_{i = 1}^{{L_r} - 1} {\sum\limits_{j = i + 1}^{{L_r}} {{f_1}(i,j,k,q){{\left( {\cos \theta _{k,i}^r - \cos \theta _{k,j}^r} \right)}^2}} } } 
            \end{aligned}
       \end{equation}
        \small
        \begin{align}
                \frac{{{\partial ^2}{\mathop{\rm g}\nolimits} }}{{\partial {x_k}\partial {y_k}}}= &\frac{{4{\pi ^2}}}{{{\lambda ^2}}}\sum\limits_{q = 1}^K {\sum\limits_{l = 1}^{{L_r}} {{f_2}} \left( l,k,q\right)\sin \theta _{k,l}^r\cos \theta _{k,l}^r\cos \phi _{k,l}^r} - \frac{{4{\pi ^2}}}{{{\lambda ^2}}}\sum\limits_{q = 1}^K {\sum\limits_{i = 1}^{{L_r} - 1} {\sum\limits_{j = i + 1}^{{L_r}} {{f_1}(i,j,k,q)\sin \theta _{k,i}^r\cos \phi _{k,i}^r\left( {\cos \theta _{k,i}^r - \cos \theta _{k,j}^r} \right)} } }\nonumber \\
                +&\frac{{4{\pi ^2}}}{{{\lambda ^2}}}\sum\limits_{q = 1}^K {\sum\limits_{i = 1}^{{L_r} - 1} {\sum\limits_{j = i + 1}^{{L_r}} {{f_1}(i,j,k,q)\sin \theta _{k,j}^r\cos \phi _{k,j}^r\left( {\cos \theta _{k,i}^r - \cos \theta _{k,j}^r} \right)} } } 
     \end{align} 
 \setcounter{equation}{\value{my1}}
	\hrulefill
	\vspace*{-6pt}
\end{figure*}

1) Subproblem with respect to $\left\{ {{{\bf{w}}_k}} \right\}$: For any given variables $\left\{ {{{\bf{u}}_k}} \right\}$, $\left\{ {{t_{k,q}}} \right\}$, (P2) can be transformed into (P3).
\begin{equation}
    \begin{aligned}
        \left({{\textrm{P3}}}\right)\mathop {\min }\limits_{\left\{ {{{\bf{w}}_k}} \right\}} \sum\limits_{k = 1}^K {{{\left\| {{{\bf{w}}_k}} \right\|}^2}}  + \frac{1}{{2\rho }}\left( {\sum\limits_{k = 1}^K {\sum\limits_{q = 1}^K {{{\left| {{{\bf{h}}_k}{{({{\bf{u}}_k})}^H}{{\bf{w}}_q} - {t_{k,q}}} \right|}^2}} } } \right).
    \end{aligned}
\end{equation}
Since this is an unconstrained quadratic convex problem, we can obtain the optimal transmit beamforming vector ${\bf{w}}_k$ by equating the first-order partial derivative of the objective function with respect to ${\bf{w}}_k$ to zero, which has the closed-form solution as following
\begin{equation}
	\small
    \begin{aligned}
        {\bf{w}}_k^ *  = \frac{1}{{2\rho }}{{\bf{A}}^{ - 1}}\left( {\sum\limits_{q = 1}^K {{t_{q,k}}{{\bf{h}}_q}{{\left( {{u_q}} \right)}^ * }} } \right){\rm{ }},\forall k,
    \end{aligned}
\end{equation}
where ${\bf{A}} = {{\bf{I}}_N} + \frac{1}{{2\rho }}\sum\limits_{q = 1}^K {{{\bf{h}}_q}({{\bf{u}}_q})} {{\bf{h}}_q}{({{\bf{u}}_q})^H}$. Since all ${\bf{w}}_k$ for different users in the objective function are separated from each other, they can be updated simultaneously according to the above equation.

2) Subproblem with $\left\{ {{t_{k,q}}} \right\},\forall k,q$: For any given beamforming vector $\left\{{{\bf{w}}_k} \right\}$ and the FA’s position $\left\{ {{{\bf{u}}_k}}\right\}$, we solve (P2) with constraint (7a) to optimize auxiliary variables ${\left\{ {{t_{k,q}}} \right\}}$. It is not difficult to observe that the auxiliary variables of different users are separated from each other, so the resulting problem can be split into $K$ independent parallel subproblems to be solved at the same time. In particular, by ignoring constant items, the corresponding subproblem for user $k$ can be simplified to
\begin{subequations}
    \begin{align}
        \left({{\textrm{P4}}}\right){\rm{}~~}&\mathop {\min }\limits_{\left\{ {{t_{k,q}},\forall q} \right\}} {\rm{~~}}\sum\limits_{q = 1}^K {{{\left| {{{\bar t}_{k,q}} - {t_{k,q}}} \right|}^2}} ,\nonumber\\
        \rm{s.t.}\quad&\frac{{{{\left| {{t_{k,k}}} \right|}^2}}}{{\sum\limits_{q = 1.q \ne k}^K {{{\left| {{t_{k,q}}} \right|}^2} + {\sigma ^2}} }} \ge {\gamma _k},
    \end{align}
\end{subequations}
where ${\bar t_{k,q}} = {{\bf{h}}_k}{({{\bf{u}}_k})^H}{{\bf{w}}_q}$, $k,q = 1,2, \cdots , K$. (P4) is a problem with quadratic objective and one quadratic inequality constraint. Although the constraint (11a) is non-convex,\cite{Boyd2004} has demonstrated that such form problem exhibit strong duality, provided Slater’s condition holds, i.e., (P4) is strictly feasible, which implies that the optimal solution can be determined through the Lagrangian duality method. Let ${\lambda _k} \ge 0,\forall k $ represent the dual variables, then the Lagrangian duality function corresponding to (P4) is expressed as follows
\begin{align}
	\small
        {\cal L}\left( {{\lambda _k},\left\{ {{t_{k,q}}} \right\}} \right)&=\left( {1 - {\lambda _k}} \right){\left| {{t_{k,k}}}\right|^2}+\sum\limits_{q = 1,q\ne k}^K{\left({1 +{\lambda _k}{\gamma _k}}\right){{\left|{{t_{k,q}}} \right|}^2}} , \nonumber \\
        & - 2\sum\limits_{q = 1}^K {{\mathop{\rm Re}\nolimits} \left\{ {{{\bar t}_{k,q}}t_{k,q}^H} \right\}} .
\end{align}
Thus, the associated dual function can be written as ${\cal G}({\lambda _k}) = \mathop {\inf }\limits_{\left\{ {{t_{k,q}}} \right\}} {\cal L}\left( {{\lambda _k},\left\{ {{t_{k,q}}} \right\}} \right)$. By setting the first-order partial derivative of the Lagrangian function in (12) with respect to $\lambda_k$ to zero, we can derive the optimal solution that minimizes ${\cal L}\left( {{\lambda _k},\left\{ {{t_{k,q}}} \right\}} \right)$ as following
\begin{equation}
    t_{k,k}^* = \frac{{\bar{t}_{k,k}}}{{1 - \lambda_k}}\quad ,\quad
    t_{k,q}^* = \frac{{\bar{t}_{k,q}}}{{1 + \lambda_k \gamma_k}}, \quad q \ne k.
\end{equation}
When the constraint in (11a) satisfies the equation, by substituting the derived $t_{k,k}^ * $ , $t_{k,q}^ * $  back (11a), this equality constraint can be transformed into another form as follows
\begin{equation}
    {\cal F}\left( {{\lambda _k}} \right) = \frac{{{{\left| {{{\bar t}_{k,k}}} \right|}^2}}}{{{{\left( {1 - {\lambda _k}} \right)}^2}}} - {\gamma _k}\sum\limits_{q = 1,q \ne k}^K {\frac{{{{\left| {{{\bar t}_{k,q}}} \right|}^2}}}{{{{\left( {1 + {\lambda _k}{\gamma _k}} \right)}^2}}}}  - {\gamma _k}{\sigma ^2} = 0.
\end{equation}
It is easy to observe that  ${\cal F}\left( {{\lambda _k}} \right)$ is a monotonically increasing function of $\lambda_k$ in the region $0 \le {\lambda _k} < 1$, so we can obtain the optimal duality variable by the bisection search\cite{wu2020joint}. However, if the constraint equation in (11a) doesn't hold, i.e., $\lambda_k=0$, Eq. (13) degenerate to  $t_{k,q}^* = \bar t_{k,q}$, $\forall k,q$.

3) Subproblem with $\left\{ {{{\bf{u}}_k}} \right\}$ : For any given variables $\left\{ {{{\bf{w}}_k}} \right\}$, $\left\{ {{t_{k,q}}} \right\}$, the antenna position $\left\{ {{{\bf{u}}_k}} \right\}$ can be optimized by solving (P2) with constraint ${\textrm{(7c)}}$. Given that the position variables for each user $\left\{ {{{\bf{u}}_k}} \right\}$ are mutually independent, (P5) can be decomposed into parallel subproblems. For user $k$, the respective subproblem is as follows:
\begin{subequations}
    \begin{align}
        \left({{\textrm{P5}}}\right){\rm{~~}}&\mathop {\min }\limits_{{{\bf{u}}_k}} {\rm{  }}\sum\limits_{q = 1}^K {{{\left| {{{\bf{h}}_k}{{\left( {{{\bf{u}}_k}} \right)}^H}{{\bf{w}}_q} - {t_{k,q}}} \right|}^2}},\nonumber\\
        \rm{s.t.}\quad&{{\bf{u}}_k} \in {{\cal C}_k},\forall k.
    \end{align}
\end{subequations}
Since ${\left| {{{\bf{h}}_k}{{\left( {{{\bf{u}}_k}} \right)}^H}{{\bf{w}}_q}-{t_{k,q}}} \right|^2}={{\bf{h}}_k}{\left( {{{\bf{u}}_k}} \right)^H}{{\bf{w}}_q}{\bf{w}}_q^H{{\bf{h}}_k}\left( {{{\bf{u}}_k}} \right) - 2{\mathop{\rm Re}\nolimits} \left\{ {{{\bf{h}}_k}{{\left( {{{\bf{u}}_k}} \right)}^H}{{\bf{w}}_q}t_{k,q}^ * } \right\} + {\left| {{t_{k,q}}} \right|^2}$,
the objective function in (P5) can be transformed to Eq. (16), where we define
\setcounter{equation}{24}
\begin{equation}
\begin{aligned}
        &{{\bf{B}}_k} \buildrel \Delta \over = {{\bf{\Sigma }}_k}{{\bf{G}}_k},{{\bf{C}}_{k,q}} \buildrel \Delta \over = {\bf{B}}_k^ * {{\bf{w}}_q}{\bf{w}}_q^H{\bf{B}}_k^T, {{\bf{d}}_{k,q}} \buildrel \Delta \over = {\bf{B}}_k^ * {{\bf{w}}_q},\\
        &\xi \left( {i,j,k} \right) \buildrel \Delta \over = \frac{{2\pi }}{\lambda }\left( {\rho _{k,i}^r\left( {{{\bf{u}}_k}} \right) - \rho _{k,j}^r\left( {{{\bf{u}}_k}} \right)} \right),\\
        &{f_1}(i,j,k,q) \buildrel \Delta \over = 2\left| {{{\bf{C}}_{k,q}}\left( {i,j} \right)} \right|\cos \left( {\xi (i,j,k) + \angle {{\bf{C}}_{k,q}}(i,j)} \right) ,\\
        &{f_2}(l,k,q) \buildrel \Delta \over = 2\left| {{t_{k,q}}} \right|\left| {d_{k,q}^l} \right|\cos \left( {\frac{{2\pi }}{\lambda }\rho _{k,l}^r\left( {{{\bf{u}}_k}} \right) + \angle d_{k,q}^l - \angle {t_{k,q}}} \right),\nonumber
\end{aligned}
\end{equation}
and ${d_{k,q}^l}$ as the $l$-th element in the vector ${{\bf{d}}_{k,q}}$.   
To tackle the resulting non-convex problem, the successive convex approximation (SCA) method is employed to optimize the position of the FA for the $k$-th user. Drawing upon Taylor’s theorem, we can construct a quadratic surrogate function that serves as a global upper bound for the objective function. Given the provided local point ${{\bf{u}}_k^i}$ in the $i$-th iteration, an upper bound can be obtained as  ${\mathop{\rm g}\nolimits} ({{\bf{u}}_k}) \le {\mathop{\rm g}\nolimits} \left( {{\bf{u}}_k^i} \right) + \nabla {\mathop{\rm g}\nolimits} {\left( {{\bf{u}}_k^i} \right)^T}\left( {{{\bf{u}}_k} - {\bf{u}}_k^i} \right) + \frac{{{\delta _k}}}{2}{\left( {{{\bf{u}}_k} - {\bf{u}}_k^i} \right)^T}\left( {{{\bf{u}}_k} - {\bf{u}}_k^i} \right)$, which is achieved through the introduction of a positive real number ${\delta _k}$ such that ${\delta _k}{{\mathbf{I}}_2} \succeq {\nabla ^2}\operatorname{g} \left( {{{\mathbf{u}}_k}} \right)$. Since
$\left\| {{\nabla ^2}g\left( {{{\bf{u}}_k}} \right)} \right\|_2^2 \le \left\| {{\nabla ^2}g\left( {{{\bf{u}}_k}} \right)} \right\|_F^2= {\left( {\frac{{{\partial ^2}{\mathop{\rm g}\nolimits} }}{{\partial {x_k}^2}}} \right)^2} + {\left( {\frac{{{\partial ^2}{\mathop{\rm g}\nolimits} }}{{\partial {x_k}\partial {y_k}}}} \right)^2} + {\left( {\frac{{{\partial ^2}{\mathop{\rm g}\nolimits} }}{{\partial {y_k}^2}}} \right)^2} + {\left( {\frac{{{\partial ^2}{\mathop{\rm g}\nolimits} }}{{\partial {y_k}\partial {x_k}}}} \right)^2}$, the $\delta_k$ can be selected by calculating the Frobenius norm of the Hessian matrix of ${\mathop{\rm g}\nolimits} ({{\bf{u}}_k})$.
According to Eq. (17)-Eq. (19), we obtain the $\delta_k$ as following:
\setcounter{equation}{19}
\begin{equation}
	\small
 {\delta _k} = \frac{{16{\pi ^2}}}{{{\lambda ^2}}}\sum\limits_{q = 1}^K {\left( {\sum\limits_{l = 1}^{{L_r}} {\left| {{t_{k,q}}} \right|\left| {d_{k,q}^l} \right|}  + \sum\limits_{i = 1}^{{L_r} - 1} {\sum\limits_{j = 1}^{{L_r}} {\left| {{{\bf{C}}_{k,q}}(i,j)} \right|} } } \right)}.
\end{equation}
In this way, (P5) is reduced to (P6).
\begin{subequations}
	\small
	    \begin{align}
        \left({{\textrm{P6}}}\right){\rm{~~}}\mathop {\min }\limits_{{{\mathbf{u}}_k}} {\text{  }}&\operatorname{g} \left( {{\mathbf{u}}_k^i} \right) + \nabla \operatorname{g} {\left( {{\mathbf{u}}_k^i} \right)^T}\left( {{{\mathbf{u}}_k} - {\mathbf{u}}_k^i} \right)\nonumber\\
       & + \frac{{{\delta _k}}}{2}{\left( {{{\mathbf{u}}_k} - {\mathbf{u}}_k^i} \right)^T}\left( {{{\mathbf{u}}_k} - {\mathbf{u}}_k^i} \right),\nonumber \\
        \rm{s.t.}\quad&{{\bf{u}}_k} \in {{\cal C}_k}.
    \end{align}
\end{subequations}
Since the objective function of (P6) is a quadratic convex function with respect to ${{\bf{u}}_k}$, ignoring the constraints leads to a closed-form solution with the global minimum solution ${\bf{u}}_k^ \star  = {\bf{u}}_k^i - \frac{{\nabla {\mathop{\rm g}\nolimits} \left( {{\bf{u}}_k^i} \right)}}{{{\delta _k}}}$. If 
${\bf{u}}_k^\star$ satisfies the constraints in (P6), then it is the optimal solution to problem (P6). Otherwise, since the antenna moving region ${{\cal C}_k}$ is a linear region, thus the problem is a quadratic programming (QP) problem and a local solution can be obtained by using quadprog.

\subsection{Update Penalty Coefficient}
Considering that the equality constraints in (P1) should be hold when our proposed algorithm converged. Therefore, the penalty cofficient is gradually decreased as following $\rho : = c\rho $, $0 < c < 1$, where $c$ is a  constant scaling factor.

Drawing upon the derivations presented above, the proposed algorithm for joint antenna position and beamforming design is summarized in $\textbf{Algorithm  1}$ straightforwardly. In the inner laye, the optimization variables $\left\{ {{{\bf{w}}_k}} \right\}$, $\left\{ {t_{k,q}} \right\}$ and $\left\{ {{{\bf{u}}_k}} \right\}$ are alternately updated until convergence of the objective function is achieved. In the outer layer, an indicator $\xi$ is defined as $\xi  = \max \left\{ {{{\left| {{{\bf{h}}_k}{{\left( {{{\bf{u}}_k}} \right)}^H}{{\bf{w}}_q} - {x_{k,q}}} \right|}^2},\forall k,q} \right\}$ and the penalty cofficient $\rho$ is updated untill $\xi$ is below a threshold. 

\subsection{Algorithm Complexity Analysis}
In the following, we analyze the computational complexity of $\textbf{Algorithm  1}$. To be specific, the complexity for solving (P3) is ${\cal O}\left( {\left( {{L_r}{L_t} + {L_t}N + {N^2}} \right)K + {N^3}} \right)$, that of solving (P4) is ${\cal O}\left( {{K^2}{{\log }_2}\left( {1/{\varepsilon _3}} \right)} \right)$ where $\varepsilon _3$ is the accuracy for the bisection search, that of solving (P6) is ${\cal O}\left( {KL_r^2{\varsigma _1} + {K^{1.5}}\ln \left( {1/\beta } \right){\varsigma _2}} \right)$ where $\beta$ is the accuracy for the interior-point method, $\varsigma _1$ and $\varsigma _2$ denote the maximum number of iterations to perform SCA and QP respectively. Thus, the overall complexity of Algorithm 1 is ${\cal O} \big( {I_{inn}}{I_{out}} \big( \left( {{L_r}{L_t} + {L_t}N + {N^2}} \right)K + {N^3} + {K^2}{{\log }_2}\left( {1/{\varepsilon _3}} \right) + KL_r^2{\varsigma _1} + {K^{1.5}}\ln \left( {1/\beta } \right){\varsigma _2} \big) \big) $ where ${I_{inn}}$ and ${I_{out}}$ denote the inner and outer iteration numbers required for convergence.  
\begin{small}
\begin{algorithm}[H]
		\caption{Joint antenna position and beamforming design algorithm}
		\begin{algorithmic}[1]
			\REPEAT
			\REPEAT
			\STATE Obtain the transmit precoders by solving (P3).
			\STATE Obtain the auxiliary variables by solving (P4).
			\STATE Obtain the optimal antenna position by solving (P6).
			\UNTIL The fractional decrease of the objective value is below a threshold ${\varepsilon _1}$.
			\STATE Update the penalty cofficient $\rho$ .
			\UNTIL The constraint violation $\xi $ is below a threshold ${\varepsilon _2} $.
		\end{algorithmic}
\end{algorithm}
\end{small}

\section{Simulation Results}
\begin{figure*}[htbp]
	\centering
	\begin{minipage}[t]{0.32\textwidth}
		\centering
		\includegraphics[width=1\textwidth]{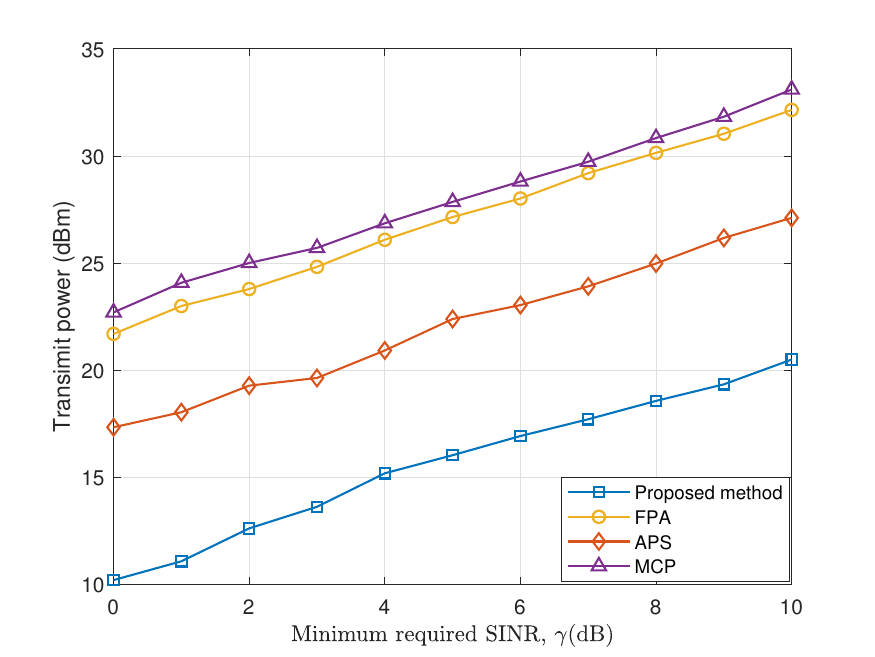}
		\setlength{\abovecaptionskip}{-0.6cm}
		\caption{Transmit power versus the minimum required SINR}
	\end{minipage}
	\begin{minipage}[t]{0.32\textwidth}
		\centering
		\includegraphics[width=1\textwidth]{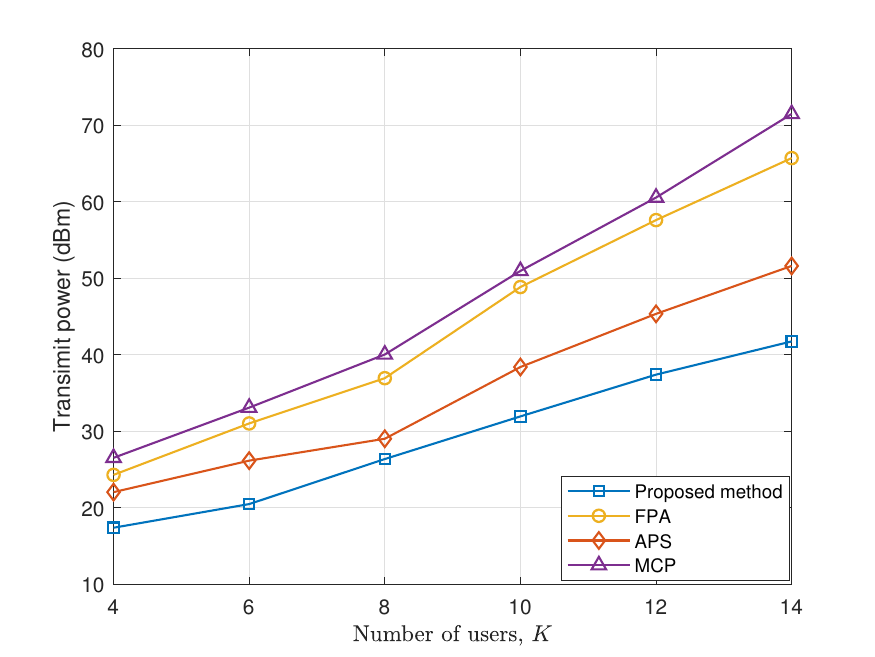}
		\setlength{\abovecaptionskip}{-0.6cm} 
		\caption{Transmit power versus the number of users.}
	\end{minipage}
	\begin{minipage}[t]{0.32\textwidth}
		\centering
		\includegraphics[width=1\textwidth]{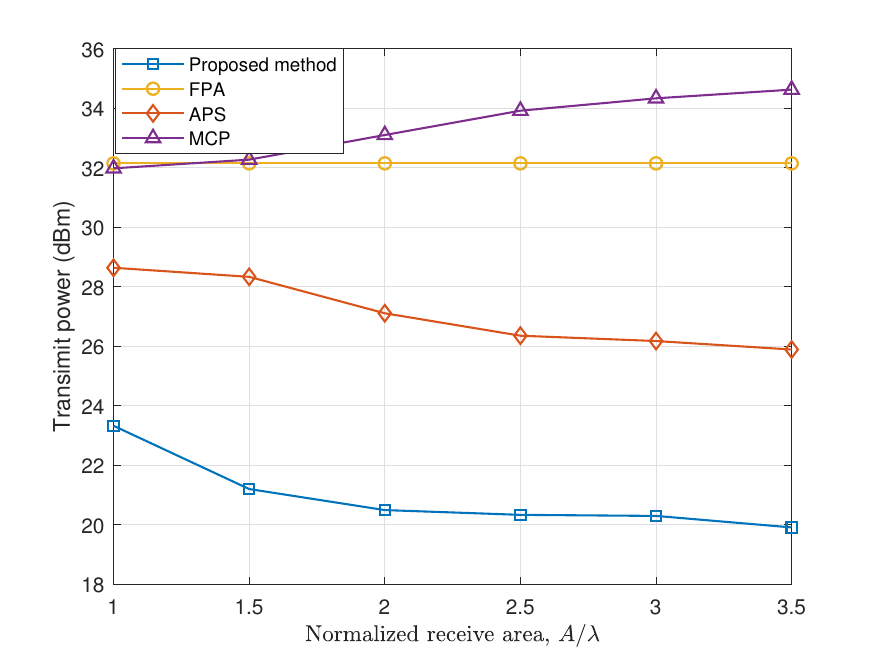}
		\setlength{\abovecaptionskip}{-0.6cm}
		\caption{Transmit power versus the normalized region size for FA.}
	\end{minipage}
\vspace{-0.6cm}
\end{figure*}
In the simulations, we contemplate a scenario in which a BS with  FPA array comprising $N = 16$ elements serves $K$ users, each of whom is equipped with a single FA. The distance between user $k$ and the BS is assumed to be a random variable following uniform distributions from 20 to 100 meters (m). The moving region for FAs at each user is set as a square area of size $A \times A$, and the initial positions are all set at the origin points, i.e., $\left\{ {{\bf{u}}_k^0} \right\}_{k = 1}^K = {[0,0]^T}$. Assuming that all users have the same number of transmit and receive paths, i.e., $L_k^t=L_k^r=10$, $1 \le k \le K$. Set the PRM of each user is a diagonal matrix with each diagonal element following distribution ${\cal C}{\cal N}\left( {0,{c_0}d_k^{ - \alpha }/L} \right)$, where ${c_0}=-40$ dB denotes the expected value of the average channel gain at the reference distance of 1 m, and $\alpha=2.8$ represents the path-loss exponent. The noise variance of each user is set to $-80$ dBm.The elevation and azimuth AoAs/AoDs are assumed to be independent and identically distributed (i.i.d.) variables following the uniform distribution over $\left[ {-\frac{\pi }{2},\frac{\pi }{2}} \right]$. The scaling factor $c$ for updating penalty coefficient is set as 0.9.

We consider three baseline schemes for comparison. (1) FPA: the antenna of each user is fixed at the origin of its local coordinate system. (2) Alternating position selection (APS): the receive area is quantized into discrete locations with equal distance $D = \lambda /2$. (3) Maximum channnel power (MCP): the MA of each user is deployed at the position which maximizes its channel power.

Fig. 2 shows the total BS transmit power required for the considered schemes versus the users’ minimum required SINR values. Under the assumption $K=6$ and $A=2\lambda$,  it is observed that the power increases as the SINR threshold increases, which means the BS needs more transmit power to meet the more stringent user quality of service requirements. Meanwhile, when the SINR is the same, our proposed algorithm can save more power than the three baseline schemes.

Fig. 3 studies the relationship between the transmit power and the number of users in different schemes under the setup $A=2\lambda$, SINR=10 dB. It can be seen that as the number of users increases, the transmit power also increases for all schemes. This is  attribute to the fact that the BS needs to consume more energy to mitigate the mutual interference between users. In addition, when the number of users increases, our proposed algorithm has better gain and the gap between the MA and FA and MCP schemes become larger. 
 
Finally, Fig. 4 describes the change of transmit power with the normalized sizes for moving region under the setup $K=6$, SINR=10 dB. It can be seen that as the size increases, our proposed algorithm decreases rapidly and stabilizes when $A = 2\lambda$. This is because that larger moving region brings more flexibility of FAs to reconfigure the channel, so that users can receive a stronger signal. In addition, the stabilisation at $A = 2\lambda$ implies that significant performance improvements can be achieved even when the antenna is moved within a relatively small area.
\section{Concluision}
In this paper, we investigate joint antenna location and beamforming design for FA-assisted  multi-user downlink communication system, in which multiple  single FA users are served by a BS equipped with a FPA array. First, we characterize the multi-user downlink channel  with respect to the position of each user's FA. Then, we formulate an optimization problem to minimize the BS transmit power under the constraints of the  minimum SINR and limited antenna moving regions. To solve this non-convex problem, an AO algorithm based on penalty method and SCA was developed to obtain a suboptimal solution. Numerical simulation results validate the proposed algorithm for FA-assisted downlink multi-user communication systems can save more power compared to FPA based systems. 
\ifCLASSOPTIONcaptionsoff
  \newpage
\fi

\bibliographystyle{IEEEtran}
\bibliography{main}
\end{document}